\documentclass{article}
\usepackage{amssymb}
\usepackage{amsfonts}
\usepackage{amsmath}

\input{tcilatex}

\begin{document}

\title{Self Creation Cosmology An Alternative Gravitational Theory}
\author{Garth A Barber}
\maketitle

\begin{abstract}
A question is raised about the premature acceptance of the standard
cosmological model, the {'}$\Lambda$CDM{'} paradigm; the non-metric, or
semi-metric, theory of Self Creation Cosmology is offered as an alternative
and shown to be as equally concordant as the standard model with observed
cosmological constraints and local observations. In self-creation the Brans
Dicke theory is modified to enable the creation of matter and energy out of
the self contained gravitational and scalar fields; such creation is
constrained by the local conservation of energy so that rest masses vary
whereas the observed Newtonian Gravitation {'}constant{'} does not. As a
consequence there is a conformal equivalence between self-creation and
General Relativity \textit{in vacuo}, which results in the predictions of
the two theories being equal in the standard tests. In self-creation test
particles \textit{in vacuo} follow the geodesics of General Relativity.
Nevertheless there are three types of experiment that are able to
distinguish between the two theories. There are also other local and
cosmological observations that are readily explained by self-creation, such
as the anomalous sunwards acceleration of the Pioneer spacecraft and a
secular spinning up of the Earth's rotation that both {'}coincidentally{'}
echo Hubble{'}s constant. Moreover, the most significant feature of
self-creation is that it is as consistent with cosmological constraints in
the distant supernovae data, the Cosmic Microwave Background anisotropies
and primordial nucleo-synthesis, as the standard paradigm. Unlike that
model, however, it does not require the addition of the undiscovered physics
of Inflation, dark non-baryonic matter, or dark energy. Nevertheless it does
demand an exotic equation of state, which requires the presence of false
vacuum energy at a moderate density determined by the field equations.
Consequently it is able to interface gravitation and quantum theories
without creating a {'}Lambda{'} problem. In self-creation there are two
frames of interpretation of observational data, which depend on whether
energy or energy-momentum is to be conserved and whether photons or atoms
respectively are chosen as the invariant standards of measurement. In the
former frame the universe is stationary and eternal with exponentially
shrinking rulers and accelerating atomic clocks, and in the latter frame the
universe is freely coasting, expanding linearly from a Big Bang with rigid
rulers and regular atomic clocks. A novel representation of space-time
geometry is suggested. As the theory is readily falsifiable it is
recommended that all three of the definitive experiments be performed at the
earliest opportunity.
\end{abstract}

\begin{center}
\textit{Note}
\end{center}

The speed of light \textit{in vacuo} is unity in this paper, except when
predictions are to be compared to observation, when the speed of light will
be explicitly designated as c.

\section{Introduction}

The present standard cosmology based on General Relativity (GR) is known as
the $\Lambda$CDM model, where {'}$\Lambda${'} refers to the dark energy, or
the cosmological constant, and {'}CDM{'} to the cold dark matter that are
believed to pervade the universe. The general consensus, [1], [2], is that
the detailed observational verification of this $\Lambda$CDM model, in which
the mass of the universe consists of 23 per cent dark matter, 73 per cent
dark energy and just 4 per cent ordinary matter, has been robustly
established and there is little purpose in exploring possible alternatives.
However, it is always healthy for the scientific process to have heterodox
theories against which the standard theory may be tested. The problem in
cosmology has been to find such a theory that not only matches General
Relativity in all the standard observations, but also raises questions that
lead to further experiments against which both theories may be tested. One
possible alternative, first published in 2002, is described in this paper;
it is the theory of Self Creation Cosmology (SCC) [3].

\subsection{Questions raised by the Standard Model}

The $\Lambda$CDM model has gained strong support from the peaks in the WMAP
and Boomerang CMB data as well as the S/N Ia Hubble diagram. As such, this
recent period has been hailed as the {'}era of precision cosmology{'} with
the general acceptance that the particular values of the cosmological
constraints mentioned above have been established beyond reasonable doubt.

Let us however examine the present problems that have been identified with
the theory. The hypothesis of Inflation was proposed in the 1970{'}s in
order to escape the density, smoothness and horizon problems of GR
cosmology; this led to the present $\Lambda$CDM model. This model suggests
that the false vacuum energy on which Inflation depended has not entirely
disappeared but continues as a remnant to constitute a significant
cosmological component called dark energy. The existence of such a false
vacuum energy is consistent with cosmic acceleration; nevertheless it is
difficult to explain why it seems to have been fine-tuned to at least one
part in $10^{102}$. In addition, the standard paradigm requires the
existence of cold dark matter together with dark energy and Inflation, all
of which depend on physics as yet undiscovered by laboratory science even
after twenty years of intensive research. Inexplicably, the densities of
these {'}dark{'} entities are not only approximately equal to each other but
also roughly equal to the baryon density. Whereas it is always possible to
explain such improbable coincidences by appealing to the anthropic
principle, the explanatory power of a theory, such as Inflation, is
questionable if it can only account for an original coincidence by the
introduction of several new ones.

There are other potential problems. In the standard $\Lambda$CDM paradigm
the SNIa Hubble diagram, mentioned above, requires a value for the vacuum
energy density that is so small that it is unstable to quantum corrections.
This vacuum energy may alternatively be interpreted as a small positive
cosmological constant; however, as such it is incompatible with the
generally accepted Superstring models that may provide the basis for a
quantum gravity theory. These theories, which compactify higher dimensions,
prefer models with a negative or zero cosmological constant.

It seems that the $\Lambda$CDM also has astrophysical problems in predicting
galaxy mass profiles that have a too pronounced cusp at small scales and a
too steep galaxy luminosity function. Although these observations are not {'}%
clean{'} in the sense that they may suffer from systematic errors that may
explain the anomalies; for example the SNIa Hubble diagram may require
evolutionary corrections and the precision of the CMB power spectrum may
still be compromised by foreground contamination from the epoch of
re-ionisation at $z = 15$; nevertheless they are also theory dependent. It
may therefore be the theoretical basis of this analysis, GR, or possibly the
cosmological topology, which requires modification.

In conclusion, the rumours of the {'}end of cosmology{'} may well be
premature.

\subsection{Other Related Anomalies?}

There are also other interesting observations closer to hand that may be
connected to cosmological effects but which have generally not been related
to fundamental physics. For example the Pioneer spacecraft appear to have an
anomalous sunward acceleration [4], [5]. This acceleration may have several
components, and there may be several possible explanations for each
component, however, as it has been observed a number of times, the excess
over the General Relativity acceleration: 
\begin{equation}
a_{P}=\left( 8.74\text{ }\pm \text{ }1.3\right) \text{ x }10^{-8}\text{
cm/sec}^{\text{2}}  \label{1}
\end{equation}

is equal to $cH$ (where $H$ is Hubble{'}s constant) if $H=87 km.sec^{-1}/Mpc$%
. Therefore this anomaly might be cosmological in nature and explained by a
non-standard gravitational theory.

A second anomaly as reviewed by Leslie Morrison and Richard Stephenson, [6],
[7], arises from the analysis of the length of the day from ancient eclipse
records. It is that in addition to the tidal contribution there is a
long-term component acting to decrease the length of the day, which equals: 
\begin{equation}
\vartriangle \text{T/day/cy}=-6\ \text{x }10^{-4}\text{ sec/day/cy.}
\label{2}
\end{equation}%
This component, which is consistent with recent measurements made by
artificial satellites, is thought to result from the decrease of the Earth{'}%
s oblateness following the last ice age. Although this explanation certainly
merits careful consideration, and it is difficult to separate the various
components of the Earth's rotation, it is remarkable that this value of $%
\vartriangle $T/day/cy is equal to $H$ if $H=67km.sec^{-1}/Mpc$. The
question is; why should this spinning up of the Earth{'}s rotation have a
natural time scale equal to the age of the universe rather than the natural
relaxation time of the order of that of the Earth{'}s crust or the
periodicity of the ice ages? This anomaly also may therefore be cosmological
rather than geophysical in nature. If this is the case then again it is a
phenomena not explicable by the standard theory.

A third anomaly, which arises from the analysis of the residues of planetary
longitudes, reveals that the Gravitational constant appears to be varying at
a rate also of the order of Hubble{'}s constant. An analysis [8] rendered a
problematic value for a variation in $G$: 
\begin{equation}
\frac{\overset{.}{G}}{G}\approx +\left( 4\pm 0.8\right) \times
10^{-11}yr^{-1}  \label{3}
\end{equation}%
with a caveat that the sign might be reversed. As this value is equal to $H$
if $H=38km.sec^{-1}/Mpc$, then it too may be cosmological in nature. GR
predicts a null result for this analysis.

If these are indeed three observations of Hubble{'}s constant, then their
values have a spread typical of other determinations of H with an average of 
$H=64km.sec^{-1}/Mpc$ in good agreement with more orthodox methods. Although
there may well be other explanations for these anomalies it is remarkable
that all three approximate Hubble{'}s constant.Although these three
observations themselves may well have non-cosmological explanations, and
individually do not seriously question GR, it will be seen that they are all
actually predicted by SCC.

\section{The Self Creation Alternative}

\subsection{Theory development 1. Mach's Principle and creation from the
Scalar Field}

The new SCC is offered as an alternative to the standard paradigm. It has
not only been shown to be concordant with cosmological observations and the
standard tests of GR, but it also provides a ready explanation for the
anomalies described above.

The SCC theories, which were first published in 1982 [9], explored
modifications of the Brans Dicke theory (BD) [10] in which the conservation
of energy-momentum was relaxed, and the equivalence principle consequently
violated, to allow mass creation. The BD theory fully incorporated Mach{'}s
principle into GR by the inclusion of an inertial scalar field while
retaining the equivalence principle. As a result, in all the SCC theories
mass creation originates from the self-contained gravitational and scalar
fields.

Although BD incorporates both the equivalence principle and Mach{'}s
principle the two might be seen to be fundamentally mutually incompatible.
Mach{'}s principle suggests that inertial frames of reference should be
coupled to the distribution of mass and energy in the universe at large.
Therefore that frame of reference in which the universe as a whole is at
rest might be considered to be a preferred {'}frame{'}, in which total
energy is conserved, in contradiction to the spirit of the equivalence
principle. Indeed such a preferred frame of reference does appear to exist;
it is that in which the Cosmic Background Radiation (CBR) is globally
isotropic. Furthermore some also argue that Mach{'}s principle requires the
gravitational parameter, G, to be determined by the large-scale structure of
the universe and not equal to some arbitrary constant.

The original SCC paper postulated two theories; the first {'}toy theory{'}
was seen to be experimentally non-concordant with the Einstein equivalence
principle (EEP), and the second was an early version of the present theory.

In order to include Mach{'}s principle the SCC theories follow BD by
coupling a scalar field that endows particles with inertia $\phi \approx 
\frac{1}{G_{N}}$ to the distribution of matter in the universe 
\begin{equation}
\Box \phi =4\pi \lambda T_{M}\text{ ,}  \label{4}
\end{equation}%
where $\Box \phi $ = $\phi _{;\;;\sigma }^{\sigma }$ is the d{'}Alembertian
of $\phi $ (the covariant equivalent of the Laplacian ${\nabla }^{2}\phi $
), $T_{M}$ is the trace of the energy-momentum tensor describing all
non-gravitational and non-scalar field energy and $\lambda $ is some
undetermined coupling constant of the order unity. The addition of the
scalar field opened up an extra degree of freedom, which is determined by $%
\lambda $. Its presence is recognised in the gravitational field equation
that now takes the form 
\begin{equation}
R_{\mu \nu }-\frac{1}{2}g_{\mu \nu }R=\frac{8\pi }{\phi }\left[ T_{M\mu \nu
}+T_{\phi \,\mu \nu }\right] \text{ ,}  \label{5}
\end{equation}%
where $T_{M\mu \nu }$ and $T_{\phi \,\mu \nu }$ are the energy momentum
tensors describing the matter and scalar fields respectively. In the second
theory the scalar field was minimally coupled to the metric and therefore
only interacted with the material universe by determining the gravitational
coefficient G hence the field equation became 
\begin{equation}
R_{\mu \nu }-\frac{1}{2}g_{\mu \nu }R=\frac{8\pi }{\phi }T_{M\mu \nu }\text{
.}  \label{6}
\end{equation}%
The 1982 paper has generated some interest (see [11]) over the last twenty
years, however Brans [12] criticised this second theory because of the
difficulty in defining the metric if photons no longer travel on
(null)-geodesics. In BD and GR photons do travel on null-geodesics because
of the equivalence principle, which leads directly to the energy-momentum
conservation equation of those theories 
\begin{equation}
\nabla _{\mu }T_{M\quad \nu }^{\quad \mu }=0\text{ .}  \label{7}
\end{equation}%
.

\subsection{Theory development 2. The Principle of Mutual Interaction}

In order to overcome Brans{'} objection that photons should follow geodesic
trajectories the latest SCC theory introduced a principle of mutual
interaction (PMI). This principle states that the scalar field is a source
for the matter-energy field if and only if the matter-energy field is a
source for the scalar field, by coupling $\nabla _{\mu }T_{M\quad \nu
}^{\quad \mu }$ to $T_{M}$, thus: 
\begin{equation}
\nabla _{\mu }T_{M\;\nu }^{.\;\mu }=f_{\nu }\left( \phi \right) \Box \phi
=4\pi f_{\nu }\left( \phi \right) T_{M\;}^{\;}\text{ ,}  \label{8}
\end{equation}%
so that for an electro-magnetic field, which is trace-free, $T_{em}=0$, 
\begin{equation}
\nabla _{\mu }T_{em\quad \nu }^{\quad \mu }=4\pi f_{\nu }\left( \phi \right)
T_{em}^{\;}=4\pi f_{\nu }\left( \phi \right) \left( 3p_{em}-\rho
_{em}\right) =0  \label{9}
\end{equation}%
thus restoring the path of a photon to a null-geodesic of the metric. As a
consequence, although SCC is not a completely metric theory, it can be
thought of as semi-metric, because photons do obey the equivalence
principle, although particles do not.

The introduction of this principle, which allows creation in a controlled
way, opened up a further degree of freedom represented by a second constant $%
\kappa $ that naturally appeared in this SCC equivalent of the conservation
field equation. Calculation, [13], determined the function $f_{\nu }\left(
\phi \right) $ above to be 
\begin{equation}
f_{\nu }\left( \phi \right) =\frac{\kappa }{8\pi }\frac{\nabla _{\nu }\phi }{%
\phi }\text{ .}  \label{10}
\end{equation}%
Consequently the conservation equation of the standard theory is replaced by
what is called the creation field equation of the new SCC theory, 
\begin{equation}
\nabla _{\mu }T_{M\;\nu }^{.\;\mu }=\frac{\kappa }{8\pi }\frac{\nabla _{\nu
}\phi }{\phi }\Box \phi \text{ .}  \label{11}
\end{equation}

This mass creation would manifest itself as a force density $f_{\nu }$ that
acts throughout space-time on non-trace-free matter according to 
\begin{equation}
f_{\nu }=\nabla _{\mu }T_{M\;\nu }^{.\;\mu }=\frac{\kappa \lambda }{2}\frac{%
\nabla _{\nu }\phi }{\phi }T_{M}\text{ .}  \label{12}
\end{equation}

The question that immediately arises is, "Is not such a violation of the
equivalence principle inconsistent with the experimental tests of the EEP
that are accurate to within one part in $10^{-14}$?"

A remarkable feature of the PMI violation of the equivalence principle is
that this {'}scalar field force{'} acts in a similar fashion to the
gravitational force. It is proportional to the product of the masses of two
freely falling bodies and inversely proportional to the square of their
separation. Thus, if this force exists, it would be convoluted with the
Newtonian gravitational force and affect the value of the Newtonian
gravitational constant in all Cavendish type experiments. In accordance with
the PMI, and as determined by the above creation field equation, this force
would affect particles but not photons. In E\"{o}tv\"{o}s-type tests of the
EEP, the details of which will be examined below, it is found that particles
of different densities fall at the same rate to within one part in $10^{-17}$
and this violation would not have been detected by EEP experiments to date.
It is, however, possible to test the theory by investigating whether photons
fall at the same rate as particles or not. GR predicts that they should
whereas SCC predicts that they should not. Thus the theory suggests a new
test of the EEP. According to SCC, the effect of the scalar field force
would be to produce two separate values for the gravitational {'}constant{'}%
, one coupled to curvature and the other measured in Cavendish type
experiments, which are {'}felt{'} by relativistic and non-relativistic
species respectively. We shall see below that the experimental effect of
this apparently gross violation of the EEP is less than might be expected
even on such projects as the LIGO gravity wave detectors.

\subsection{Theory development 3. The Local Conservation of Energy}

As in BD the presence of the scalar field perturbs space-time, affects $%
\gamma$ (the curvature produced per unit mass), and modifies the GR
geodesics of freely falling test particles. However, it was realised at an
early stage of the development of this theory that if $\kappa = 1/\lambda$
then the scalar field force had the effect of exactly compensating for this
perturbation of space-time. Subsequently it was realised that this exact
compensation was caused by the fact that this particular relationship of $%
\lambda$ and $\kappa$ created a conformal equivalence \textit{in vacuo}
between SCC and canonical GR.

Furthermore it was seen that if a body was lifted against a gravitational
field and $\kappa =1$, then the increase in rest mass would be exactly equal
to the gain in potential energy. Consequentially the theory was refined by
the introduction of another principle, that of the local conservation of
energy. Accordingly, it was postulated that a particle{'}s rest mass, $m$,
should vary with gravitational potential energy and therefore given by 
\begin{equation}
m(r)=m_{0}exp\left[ \Phi _{N}(r)\right] \text{ .}  \label{13}
\end{equation}%
Here $\Phi _{N}(r)=-\frac{G_{N}M}{rc^{2}}$, is the dimensionless Newtonian
potential and $m_{0}$ is the particle{'}s rest mass at infinity. The force
required to lift such a mass against a gravitational field is 
\begin{equation}
F_{r}=\frac{dE}{dr}=\left[ \frac{dm}{dr}\right] c^{2}=\frac{G_{N}Mm}{r^{2}}%
\text{ ,}  \label{14}
\end{equation}%
in accordance with Newtonian gravitational theory.

Mass and energy are defined at the Centre of Mass (CoM) of the gravitating
body. Whereas in BD inertial rest masses remain constant and the
gravitational {'}constant{'} $G_N$ varies, in SCC inertial masses vary and
it is the observed Newtonian constant that remains invariant. The constants $%
\lambda$ and $\kappa$ determine the density of the scalar field and the {'}%
rate{'} of creation respectively. If the local conservation of energy and
the invariance of $G_N$ are assumed then consistency requires both $\lambda$
and $\kappa$ to be unity. It was shown, [3], [13], that with these values
photons {'}fall{'} at one and a half times the gravitational acceleration
experienced by particles, thereby providing one definitive test for the
theory. As a consequence the theory is highly determined and makes specific
predictions that render it easily falsifiable. However the conformal
equivalence with canonical GR has the consequence that in all the standard
tests \textit{in vacuo} to date, the predictions of both SCC and GR are
identical. Therefore the present tests that do verify GR with precision have
not yet falsified SCC either.

In SCC, in which physical rulers and clocks vary with atomic masses, it is
light that adopts the fundamental role of measuring the universe, in a
similar, but more general, fashion to the theory of E.A.Milne{'}s Kinematic
Relativity, [14], [15].

If the above introduction of both the local conservation of energy and Mach{'%
}s principle appears contrived, it is pertinent to remember that Einstein
himself gave some consideration to these principles because he was concerned
that they were not included in GR. They have been considered independently
at various times since the publication of Einstein{'}s GR papers without
much success, but in SCC they are considered together, and then they do
produce a theory that is concordant with observation.

Nevertheless, as the conservation of energy-momentum is one of the most
fundamental principles of modern physics we may well ask for what reason
should it be abandoned? In order to formulate an answer to this question
consider the fact that energy-momentum is a manifestly covariant concept,
whereas energy is not, therefore according to the equivalence principle, and
hence GR, it is energy-momentum that is locally conserved and not energy.
This is because the total relativistic energy of a particle is relative to
the inertial frame of reference in which it is measured. As the equivalence
principle does not allow a preferred frame there is no definitive value for
energy in any metric theory in which that principle holds. In the
terminology of Emmy Noether{'}s 1918 paper [16], (see [17]), GR is an
example of an improper energy theorem. A consequence of energy-momentum
being conserved in a metric theory such as GR is that a particle{'}s rest
mass is necessarily invariant. This is a result of rest mass being
mathematically identical to the norm of the four-momentum vector.

In order to appreciate the problem with the conservation of energy-momentum
at the expense of that of energy, let us consider, in a {'}gedanken{'} or
thought experiment, the four-momentum, $P^{\mu}$, of a projectile freely
falling towards the Earth observed from an imaginary free falling laboratory
at the Earth{'}s CoM. The four-momentum vector is composed of the
mass-energy vector $P^0$ together with the momentum vector $P^i$ (where the
superscripts 0 = time and i = space). According to all metric theories the $%
P^{\mu}$ is conserved even as the projectile{'}s velocity towards the Earth
increases and hence so does its momentum relative to our observer. As $P^i$,
the momentum, increases and the norm $|P^{\mu}|$ remains constant then $P^0$%
, the total energy or {'}relativistic mass{'} of the particle cannot be
generally conserved; indeed it too steadily increases with the gain of
kinetic energy. The problem is, however, that in the GR understanding of the
situation no work is being done on or by the projectile, because it is
freely falling along a geodesic converging with the Earth{'}s geodesic, so
why should its energy increase? As a correction for this anomaly SCC
proposes that the total energy of a freely falling body, on which no work is
being done, should be invariant; a proposal that is implemented by the
definition of rest mass above.

Consequentially in general SCC violates the equivalence principle, as it is
energy and not energy-momentum that is conserved. Nevertheless the theory is
still concordant with the present tests of the EEP. As a corollary
gravitational red shift is not interpreted as a loss of energy by the photon
but as an increase in rest mass of the apparatus used to measure it, [3].
The photon itself is of invariant energy and frequency.

As the energy and frequency of a photon is thus invariant when crossing
curved space-time, a {'}standard{'} photon, suitably defined, may be used as
the unit measure for energy and hence mass, time and hence length. This
system of measurement is called the Jordan energy frame [JF(E)] of the
theory. Alternatively, a more usual system of measurement using atomic
rulers and clocks, in which atomic rest masses are invariant and which
conserves energy-momentum, may be defined; this is the Einstein frame (EF)
of the theory.

\section{The Conformal Transformation}

Weyl{'}s hypothesis [18] was that a true infinitesimal geometry could only
restrict the space-time manifold, $M$, to a class [$g_{\mu \nu }$] of
conformally equivalent Lorentz metrics and not just to a unique metric as in
GR. These metrics are related through a conformal transformation given by 
\begin{equation}
g_{\mu \nu }\rightarrow \widetilde{g}_{\mu \nu }=\Omega ^{2}g_{\mu \nu }%
\text{ .}  \label{15}
\end{equation}%
(Note: A tilde, \~{n}, signifies the Einstein frame and plain type, n, the
Jordan frame.)

Recent interest that has sought to include a scalar field with the
gravitational field has led to the development of a number of BD-type scalar
field theories. These have a Jordan frame ($g_{\mu \nu }$) wherein G varies
and particle rest masses are constant, and an Einstein Frame ($\widetilde{g}%
_{\mu \nu }$) wherein rest masses vary and it is G that is constant. However
Non Linear Gravity (NLG) theories have also been suggested that have
constant rest masses in the Einstein frame. The Lagrangian densities of
these NLG theories, the equations of which are cast in the same way as SCC,
are given by: 
\begin{equation}
L^{SCC}[g,\phi ]=\frac{\sqrt{-g}}{16\pi }\left( \phi R-\frac{\omega }{\phi }%
\phi _{;\sigma }\phi _{;}^{\sigma }\right) +L_{matter}^{SCC}[g,\phi ]
\label{16}
\end{equation}%
in the Jordan frame and 
\begin{equation}
L^{GR}[\widetilde{g},\widetilde{\phi }]=\frac{\sqrt{-\widetilde{g}}}{16\pi
G_{N}}\left[ \widetilde{R}-\left( \omega +\frac{3}{2}\right) \widetilde{g}%
^{\mu \nu }\widetilde{\nabla }_{\mu }\widetilde{\phi }\widetilde{\nabla }%
_{\nu }\widetilde{\phi }\right] +\widetilde{L}_{matter}[\widetilde{g}]
\label{17}
\end{equation}%
in the Einstein frame.

In SCC the choice of frames is one of choice of a method of measurement; the
question being, "What standard is used to define a unit of mass, length and
time and how is that standard to be transported around the universe?" In the
JF(E), in which energy is conserved, that standard is a standard photon; or
in the EF, in which energy-momentum is conserved, it is a standard mass. The
rest mass at a point $x_{%
%TCIMACRO{\U{b5}}%
%BeginExpansion
{\mu}%
%EndExpansion
}$ is defined in the JF(E) and the EF by 
\begin{equation}
m(x^{\mu })=m_{0}\exp [\Phi _{N}\left( x^{\mu }\right) ]\text{ and }%
\widetilde{m}(\widetilde{x}^{\mu })=\widetilde{m}_{0}\text{ respectively.}
\label{18}
\end{equation}%
As mass is conformally transformed according to $m\left( x^{\mu }\right)
=\Omega \widetilde{m}_{0}$, this duality may be obtained by defining the
conformal transformation as 
\begin{equation}
\Omega =\exp \left[ \Phi _{N}\left( x^{\mu }\right) \right] \text{ .}
\label{19}
\end{equation}%
In SCC both $\lambda $ and $\kappa $ are determined by the basic principles
of the theory to be unity, and as $\omega $ is given by 
\begin{equation}
\omega =\frac{1}{\lambda }-\frac{3}{2}-\kappa \text{ ,}  \label{20}
\end{equation}%
\begin{equation}
\text{therefore }\omega =-\frac{3}{2}\text{ .}  \label{21}
\end{equation}%
With this conformal relationship, $\Omega $, and value for $\omega $, the
conformal transformation \textit{in vacuo} of the SCC Jordan (Energy) frame 
\begin{equation}
L^{SCC}[g,\phi ]=\frac{\sqrt{-g}}{16\pi }\left( \phi R+\frac{3}{2\phi }%
g^{\mu \nu }\nabla _{\mu }\phi \nabla _{\nu }\phi \right)
+L_{matter}^{SCC}[g,\phi ]\text{ becomes}  \label{22}
\end{equation}%
\begin{equation}
L^{SCC}[\widetilde{g}]=\frac{\sqrt{-\widetilde{g}}}{16\pi G_{N}}\widetilde{R}%
+\widetilde{L}_{matter}^{SCC}[\widetilde{g}]\text{ in the Einstein frame,}
\label{23}
\end{equation}%
which is canonical GR.

This conformal equivalence with canonical GR explains why the relationship $%
\lambda = \kappa^{-1}$ leads to the exact compensation for the perturbation
of space-time by the scalar field force. Test particles follow the geodesics
of GR \textit{in vacuo}. However when not \textit{in vacuo} this conformal
equivalence breaks down and it is here that the theory differs from GR and
may be tested against it. Tests that compare in the two theories the
trajectories of geodesics \textit{in vacuo} will discover they are the same,
whereas local measurements in the two theories of curvature, gravitational
acceleration of photons, and local false vacuum densities, will differ.

In order that the reader may determine the theory{'}s predictions in
contexts other than those explicitly treated in the SCC papers, the rule to
be applied is that the curvature of space-time, and hence gravitational
orbits and cosmological equations, are to be calculated in the JF(E);
whereas atomic processes such as primordial nucleo-synthesis and the physics
of matter are best calculated in the EF.

There has been a debate about scalar field theories in general regarding
which frame is the physical frame, as both are problematic; the equivalence
principle being violated in one frame and the scalar field energy being
negative in the other. However in SCC both frames are physical, for not only
is the equivalence principle preserved in the EF but also the scalar field
energy is non-negative in the JF(E).

\section{The Centre of Mass (CoM)}

At the CoM of a system the function determining scalar field is stationary
so that $\nabla _{\nu }\phi =0$, therefore the creation equation, 
\begin{equation*}
\nabla _{\mu }T_{M\;\nu }^{.\;\mu }=\frac{\kappa }{8\pi }\frac{\nabla _{\nu
}\phi }{\phi }\Box \phi
\end{equation*}%
becomes 
\begin{equation}
\nabla _{\mu }T_{M\;\nu }^{.\;\mu }=0  \label{25}
\end{equation}%
and the conservation equation is regained. Hence, at the unique location of
the CoM of the system the energy-momentum tensor of matter is conserved with
respect to covariant differentiation. Here the theory admits a ground state
solution, the metric tensor reduces to that of Special Relativity, $g_{\mu
\nu }\rightarrow \eta _{\mu \nu }$ , here the equivalence principle holds,
even for a massive particle, and here a {'}freely falling{'} physical clock
records proper time and standards of mass, length, time and the physical
constants, together with potential energy, retain their classical meaning.
Distances can be measured by timing the echo of light rays (radar) using the
freely falling clock and the metric may be properly defined. A {'}standard{'}
well-defined atom may emit a {'}reference{'} photon mentioned earlier with
well-defined and invariant energy, and hence frequency, which is
subsequently transmitted across space-time.

Time is the fundamental measurement in both conformal frames. It is measured
by the (invariant) frequency of the reference photon in the Jordan frame and
by the {'}Bohr{'} frequency of an atom (of invariant rest mass) in the
Einstein frame. The speed of light is invariant in both. It is contended
that gravitational, and hence cosmological, problems have to be solved in
the Jordan frame and this is used throughout unless specifically stated
otherwise.

The CoM is therefore selected as a preferred frame of reference by the
requirement to locally conserve energy in the Jordan Frame. It divides out
an {'}absolute{'} time from the manifold thus selecting a "preferred
foliation of space-time", to use Butterfield and Isham{'}s expression [19],
and such a preferred reference frame may provide insight into the problems
at the gravitation and quantum theory interface. It might be pertinent to
investigate this question in the future, but it is not followed up in this
paper.

Over the past century physicists have wrestled with the question of absolute
reference frames and hence absolute time. Einstein reacted against those of
Lorentz and Fitzgerald by introducing the equivalence principle and the
philosophy of {'}no-preferred frames{'} of relativity theory. However
quantum gravity has struggled to reintroduce into gravitation such a
preferred foliation of space-time, for that appears to be required by
quantum considerations. In fact SCC unifies the conflicting requirements of
GR and quantum physics because it contains a manifestly covariant Einstein
frame and it also recognises a preferred foliation in its Jordan frame.
However this preferred frame is not {'}absolute{'} in the Newtonian sense of
the word but rooted by Mach{'}s principle in the distribution of mass-energy
within the universe.

\section{Gravitational Red Shift}

Consider a "gedanken", or thought, experiment in which a laboratory is
constructed at the CoM of a spherical gravitating body connected to the
surface by a radial tunnel through which photons and test particles may be
projected {'}\textit{in vacuo}{'} to various maximum altitudes. The initial
total energy or relative mass of each projectile as it is launched consists
of its rest mass plus its kinetic energy, but later when it momentarily
comes to rest at maximum altitude that kinetic energy vanishes. Thus,
according to GR, the total energy appears to decrease even though no work
has been done on or by the projectile while in freefall. The situation is
aggravated when an allowance is made for the effect of curvature. The
separation of events in space-time is described by the metric which is
defined by $d\tau ^{2}=-g_{\mu \nu }dx^{\mu }dx^{\nu }$ where the repeated
lower and upper indices indicate summation over the four dimensions. It can
be shown that if the times of arrival at ($x_{1}$) of two adjacent wave
fronts emitted from ($x_{2}$) are considered, where ($x_{1}$) and ($x_{2}$)
are mutually at rest in a gravitational field, then the emission and
absorption frequencies of photons are described by the time dilation
relationship 
\begin{equation}
\nu (x_{2})/\nu (x_{1})=\sqrt{\left[ g_{00}(x_{2})/g_{00}(x_{1})\right] }%
\text{ .}  \label{26}
\end{equation}%
Therefore a photon emitted with frequency $\nu $ at $x_{2}=r$ and received
with frequency $\nu _{0}$ at $x_{1}=\infty $, will be observed to suffer the
standard gravitational red shift of 
\begin{equation}
\nu (r)=\nu _{0}\sqrt{\left[ -g_{00}(r)\right] }\text{ .}  \label{27}
\end{equation}%
This is normally interpreted as a time dilation effect in which a clock deep
in a gravitational potential well is observed from above to run slowly when
compared with a clock at altitude, an effect that is well accepted and
observed experimentally. But if it is assumed that the energy of the
projectile in free fall is conserved, as determined in the CoM system of
coordinates, then the rest mass is given by the expression, [3], [13], 
\begin{equation}
m(r)=m_{0}exp\left[ \Phi (r)\right] \sqrt{\left[ -g_{00}(r)\right] }\text{ .}
\label{28}
\end{equation}%
In this analysis the time dilation effect applies equally to the relative
mass of a body as it does to photons projected between the same levels. Any
measurement of gravitational red shift compares the energy of a photon ($%
h\nu $) with the physical mass-energy of the atom ($m_{p}$) it interacts
with, therefore such a measurement should be described, at the CoM, by the
comparison of the two expressions above yielding: 
\begin{equation}
\frac{m(r)}{\nu (r)}=\frac{m_{0}exp\left[ \Phi (r)\right] }{\nu _{0}}\text{ .%
}  \label{29}
\end{equation}%
This relationship, which compares two observables, frequency and mass,
includes the factor accounting for a difference in potential energy but not
the one for time dilation. As these two factors are "coincidentally" equal
in GR they have been readily confused. However the pertinent question is,
"How is the above relationship to be interpreted?" That is, "How are mass
and frequency to be defined and measured at different levels in any
particular theory?" Experiments would naturally use the Einstein frame in
which atomic masses are axiomatically defined to be constant as most
apparatus is normally constructed out of atoms! 
\begin{equation}
m(r)=m_{0}\text{ .}  \label{30}
\end{equation}%
Yet the problem in this frame is that photons mysteriously lose energy
because the comparison of m(r) and n(r), yields, 
\begin{equation}
\nu (r)=\nu _{0}exp\left[ -\Phi (r)\right] \text{ .}  \label{31}
\end{equation}%
On the other hand if it is postulated that particle rest mass should include
gravitational potential energy with 
\begin{equation*}
m(r)=m_{0}exp\left[ \Phi (r)\right] 
\end{equation*}%
as in the Jordan frame of SCC, the comparison of $m(r)$ and $\nu (r)$,
yields 
\begin{equation}
\nu (r)=\nu _{0}\text{ .}  \label{33}
\end{equation}%
Thus the energy and hence frequency of a free photon is invariant in the
Jordan frame, even when it transverses curved space-time. Contrary to GR, it
might be thought self evident that as no work is done on or by a free-photon
that obeys the equivalence principle, then its energy ought to be constant.
As mentioned above, in the SCC Jordan frame gravitational red shift is
interpreted not as a loss of potential energy by the photon but rather as a
gain of potential energy, and therefore mass, by the observer{'}s apparatus.

The problem with the equivalence principle can now be restated; the rest
mass of a raised body is invariant although work has been done lifting it,
and the energy of a photon transmitted against a gravitational well deceases
even though no work is done on or by it. In the Jordan frame of SCC this
situation is reversed.

\section{The SCC field equations}

The functions $T_{\phi \,\mu \nu }$ and $f_{\nu }\left( \phi \right) $ are
calculated in the JF(E) to obtain the following set of field equations, [3]: 
\begin{equation}
\Box \phi =4\pi T_{M}^{\;}\text{ , }  \label{34}
\end{equation}%
(the scalar field equation),%
\begin{eqnarray}
R_{\mu \nu }-\frac{1}{2}g_{\mu \nu }R &=&\frac{8\pi }{\phi }T_{M\mu \nu }-%
\frac{3}{2\phi ^{2}}\left( \nabla _{\mu }\phi \nabla _{\nu }\phi -\frac{1}{2}%
g_{\mu \nu }g^{\alpha \beta }\nabla _{\alpha }\phi \nabla _{\beta }\phi
\right) \ \ \   \label{35} \\
&&+\frac{1}{\phi }\left( \nabla _{\mu }\nabla _{\nu }\phi -g_{\mu \nu }\Box
\phi \right) \text{ ,}  \notag
\end{eqnarray}%
(the gravitational field equation), 
\begin{equation}
\nabla _{\mu }T_{M\;\nu }^{\;\mu }=\frac{1}{8\pi }\frac{\nabla _{\nu }\phi }{%
\phi }\Box \phi =\frac{1}{2}\frac{\nabla _{\nu }\phi }{\phi }T_{M}\text{ , }
\label{36}
\end{equation}%
(the creation field equation),

together with some equation of state $p=\sigma \rho $ where $\sigma =+\frac{1%
}{3}$ in a radiation dominated universe and $\sigma =0$ in a dust universe.
It has been shown [3], that in the cosmological case SCC requires an exotic
equation of state $\sigma =-\frac{1}{3}$.

It is now necessary to consider the local gravitational and scalar fields
around a static, spherically symmetric, mass embedded in a cosmological
space-time. In such an embedding the value of the scalar $\phi$ defining
inertial mass asymptotically approaches a cosmological value $\frac1{G_N}$,
which holds {'}at great distance{'} from any large masses, and it is to be
defined in the inertial, preferred frame of reference, the CoM. By solving
the equations for the components of the gravitational field equation in this
restricted case, that of the Post Newtonian Approximation (PNA), the
curvatures of space-time in SCC and BD are shown to be identical in this
approximation, [3], [13]. Hence in SCC, as in BD, the gravitational field
outside a static spherically symmetric mass depends on M alone but not any
other property of the mass and the solution for the scalar field is given as

\begin{equation}
\phi =G_{N}^{-1}\exp (-\Phi _{N})\text{ .}  \label{37}
\end{equation}%
.

Also the Robertson parameters for SCC are given by the same formulas as in
BD and in SCC are: 
\begin{equation}
\alpha _{r}=1\qquad \beta _{r}=1\qquad \gamma _{r}=\frac{1}{3}\text{ .}
\label{38}
\end{equation}

Following BD, and GR, analysis of the ten PPN parameters reveals SCC to be a
conservative theory with no preferred frame effect.

As we have seen the effect of allowing 
\begin{equation*}
\nabla _{\mu }T_{M\;\nu }^{\;\mu }=\frac{1}{8\pi }\frac{\nabla _{\nu }\phi }{%
\phi }\Box \phi
\end{equation*}%
is to produce an extra {'}scalar field{'} force perturbing "free falling"
slow moving particles from their geodesic paths. This force is found to be
directly proportional to the purely gravitational force and, as the
acceleration it produces on such bodies is independent of their mass, it
behaves as Newtonian gravitation.

As a result the gravitational and scalar field accelerations would be
confused in any experiment and convoluted together. Accordingly the
Newtonian gravitational constant as measured in a Cavendish type experiment, 
$G_{N}$, is a compilation of $G_{m}$, which couples the curvature of
space-time to mass, and the effect of the scalar field. A detailed
calculation [3], yields 
\begin{equation}
G_{m}=\frac{3}{2}G_{N}\text{ .}  \label{40}
\end{equation}%
Therefore the acceleration of a massive body caused by the curvature of
space-time is 3/2 the Newtonian gravitational acceleration actually
experienced, however this is compensated by a force causing an
anti-gravitational acceleration due to the scalar field of 
%TCIMACRO{\U{bd} }%
%BeginExpansion
$\frac12$
%EndExpansion
Newtonian gravity. Note that $G_{N}$ and $G_{m}$ refer to the total
accelerations experienced in gravitational experiments by atomic particles
and photons respectively. Consequently this theory makes definite and
falsifiable predictions about observable measurements.

With these values $G_{N}$ is the proper value of $1/\phi $ at infinity, in {'%
}Cavendish{'} type experiments this value of $G_{N}$ would be measured by
atomic apparatus everywhere, as $G$ and $\phi $ only vary in the Jordan
frame (i.e. in measurements using electromagnetic and gravitational methods
alone). In SCC the Schwarzschild metric in the standard form is, 
\begin{eqnarray}
d\tau ^{2} &=&\left( 1-\frac{3G_{N}M}{rc^{2}}+..\right) dt^{2}-\frac{1}{c^2}\left( 1+%
\frac{G_{N}M}{rc^{2}}+..\right) dr^{2}  \label{41} \\
&&-\left( \frac{r}{c}\right) ^{2}d\theta ^{2}-\left( \frac{r}{c}\right)
^{2}\sin ^{2}\theta d\varphi ^{2}\text{ .}  \notag
\end{eqnarray}%
The combined effect of space-time curvature and the scalar field force is
given by the equations of motion of a freely falling particle as 
\begin{equation}
\frac{d^{2}r}{dt^{2}}=-\left\{ 1-\frac{G_{N}M}{rc^{2}}+...\right\} \frac{%
G_{N}M}{r^{2}}\text{.}  \label{42}
\end{equation}%
The effect of this non-Newtonian perturbation will be shown later.

\section{The Observational Tests of SCC}

The SCC predictions will now be examined for the three original {'}classical{%
'} tests of GR suggested by Einstein; the deflection of light by the sun,
the gravitational red shift of light and the precession of the perihelia of
the orbit of Mercury, together with three more recent tests, the time delay
of radar echoes passing the sun, the precessions of a gyroscope in earth
orbit and that {'}test-bed{'} of GR, the binary pulsar PSR 1913 + 16. In
several calculations of these tests, using the Robertson parameters, in
which $G_{m}=\frac{3}{2}G_{N}$ and $\gamma _{r}=\frac{1}{3}$, the following
factor, appears: 
\begin{equation}
\frac{\left( 1+\gamma _{r}\right) }{2}G_{m}=G_{N}\text{,}  \label{43}
\end{equation}

Thus in these cases the effect of the scalar field perturbing the curvature
of space-time from the GR curvature is exactly compensated by the effect of
the scalar field acceleration on the measurement of G. This can be seen in
the case of the deflection of light by a massive body where the deflection
is given by 
\begin{equation}
\triangle \theta =\frac{4G_{m}M}{Rc^{2}}\left( \frac{1+\gamma _{r}}{2}%
\right) =\frac{4G_{N}M}{Rc^{2}}\text{ ,}  \label{44}
\end{equation}%
which is the GR expression. SCC therefore agrees with GR and observation.
This agreement also holds for measurements of the delay in the timing of
radar echoes passing the sun and reflected off (say) Mercury, or a
spacecraft, at superior conjunction. According to Misner, et al. [20] the
deflection is given by 
\begin{equation}
\frac{d\triangle \tau }{d\tau }-(\text{Constant Newtonian part})=-8\frac{%
\left( 1+\gamma _{r}\right) }{2}\frac{G_{m}M}{bc^{2}}\frac{db}{d\tau }=-8%
\frac{G_{N}M}{bc^{2}}\frac{db}{d\tau }\text{ ,}  \label{45}
\end{equation}%
where b is the distance of the ray from the earth-sun axis. SCC thus
predicts the GR value; this has been confirmed by observations to $\pm $1
per cent. accuracy. Although the scalar field similarly compensates for the
effect of the curvature of space-time on the precession of a gyroscope in
earth orbit in the frame-dragging precession it does not do so in the
geodetic precession; so this latter measurement presents another test of the
theory, which will be discussed below. If we now examine the precession of
perihelia of an orbiting body, primarily the planet Mercury, we find that it
does not fall into this type of relationship. In terms of the Robertson
parameters the precession is given by the expression 
\begin{equation}
\triangle \theta =\frac{6\pi G_{m}M}{Lc^{2}}\frac{\left( 2-\beta
_{r}+2\gamma _{r}\right) }{3}\text{ radians/orbit.}  \label{46}
\end{equation}%
Whereas in GR 
\begin{equation}
\triangle \theta =\frac{6\pi G_{N}M}{Lc^{2}}  \label{47}
\end{equation}%
in agreement with observation to an accuracy $\pm $1 per cent. Substituting
the relevant values of $\gamma _{r}$ and $G_{m}$ for SCC in the above
expression yields 
\begin{equation}
\triangle \theta =\frac{5\pi G_{N}M}{Lc^{2}}\text{ .}  \label{48}
\end{equation}%
However as the planet is a massive body it is subject to the scalar field
acceleration, which modifies the Newtonian gravitation field with a
dipole-like potential, so the total potential is 
\begin{equation}
\Phi =-\frac{G_{N}M}{r}+\frac{1}{2}\left( \frac{G_{N}M}{rc}\right) ^{2}
\label{49}
\end{equation}

This is the well-known "semi-relativistic" potential obtained by enhancing
the mass of an orbiting body with its potential energy and which produces a
precession of 
\begin{equation}
\triangle \theta {^{\prime }}=\frac{\pi G_{m}M}{Lc^{2}}\text{ radians/orbit.}
\label{50}
\end{equation}

Therefore, if one combines the precession calculated from the metric in SCC
with the perturbation caused by the scalar field force, one obtains a
prediction of precession for SCC that is exactly the same as GR, 
\begin{equation}
\triangle \theta =\frac{6\pi G_{m}M}{Lc^{2}}\text{ radians/orbit.}
\label{51}
\end{equation}

In SCC a neutron star, composed of relativistic matter with an equation of
state of $p=\frac{\rho c^2}3$, will be decoupled from the scalar field. Any
predictions about the loss of orbital energy by the binary pulsar PSR1913+16
due to gravitational radiation would appear be the same as GR. However in
the process of formation of such a collapsed star its gravitational field
would appear to increase by a factor of 3/2 as its matter became degenerate
and decoupled from the scalar field. This would assist the gravitational
collapse of such an object so that the minimum mass limit, the Chandrasekhar
limit, for a completely degenerate core is reduced from 1.4 to 0.93 solar
mass, although this difference in mass would not be detectable except in the
transitional case of a binary system caught in the act of becoming
degenerate.

\subsection{Experimental Tests of the Theory}

It may be difficult for some to believe that a theory so different from GR
could predict the same outcomes in all previous standard tests. It is
therefore important to recall that \textit{in vacuo} SCC test particles
follow the trajectories of GR geodesics. Consistent with this is the fact
that in SCC although the Robertson parameter $\gamma = \frac 13$, whereas in
GR $\gamma = 1$; this is compensated in most observations by an increase of $%
G_m$ of 50 per cent. Consequently, we have been able to show that SCC is
concordant with all those experiments that otherwise have been thought to
verify GR.

However, may not the above identical predictions of GR and SCC in the
One-Body Problem raise the suspicion that SCC is just GR rewritten in some
obscure coordinate system? The existence of at least three types of
experimental test that are suggested by SCC proves that this is not so. In
this section we shall examine these definitive tests and in the next we
shall consider the particular problem of the EEP.

The first type of test poses the following question; "Do photons and
particles fall {'}at the same rate{'}?" The prediction of the deflection of
light by massive bodies is equal in both theories when observed at a
distance; this is actually a scattering experiment. Nevertheless SCC
predicts that a photon in free fall descends at 3/2 the acceleration of
matter, i.e. in free fall a beam of light travelling a distance l is
deflected downwards, relative to physical apparatus, by an amount 
\begin{equation}
\delta =\frac{1}{4}g_{Earth}\left( \frac{l}{c}\right) ^{2}\text{ ,}
\label{52}
\end{equation}%
where $g_{Earth}$ is the terrestrial Newtonian gravitational acceleration.
As a possible space experiment I suggest an annulus, two meters in diameter
supporting, for example, 1,000 carefully aligned small mirrors. A laser beam
is then split, one half reflected, say 1,000 times, to be returned and
recombined with the other half beam, reflected just once, to form an
interferometer at source. If the experiment is in earth orbit and the
annulus orientated on a fixed star, initially orthogonal to the orbital
plane then the gravitational or acceleration stresses on the frame would
vanish, whereas they would predominate on earth. In low Earth orbit SCC
predicts a 2 \AA ngstrom ($2\times 10^{-10}m$) interference pattern shift
with a periodicity equal to the orbital period whereas GR predicts a null
result.

This test may be carried out on Earth using the fact that the Earth is in
free fall around the Sun and is accelerating radially towards it at about $%
0.01m/sec^{2}$. For example, the laser beams of the LIGO gravity wave
telescope travel horizontally along two orthogonal 4km tunnels and are then
reflected back to be re-combined at an interferometer at source. For the
experiment one beam could be returned immediately by an additional mirror to
give it a negligible path length and then compared with the other beam that
had travelled 8 km. The theory predicts that the two beams would then be
displaced relative to each other in a direction towards the Sun by an amount 
\begin{equation}
\delta =\frac{1}{4}g_{Sun}\left( \frac{l}{c}\right) ^{2}\approx 2\times
10^{-12}\text{metres ,}  \label{53}
\end{equation}%
where $g_{Sun}$ is the Newtonian gravitational acceleration of the Earth
towards the Sun.

Although the LIGO interferometers can measure a longitudinal displacement to
an accuracy of the order $10^{-18}m.$, whether they could have already
detected a diurnal beam displacement of $2\times 10^{-12}m.$, normal to the
line of sight, is an interesting question. Nevertheless the adaptation of
this existing apparatus with a suitably constructed interferometer may well
be the cheaper method of testing whether light does {'}fall{'} at the same
rate as matter, or otherwise.

The second type of test poses the following question; "Is there a limit to
the Casimir force that is dependent on space-time curvature?" This question
arises from the real vacuum solutions of the scalar potential, [3], which
yield small non-zero densities if the curvature is non-zero. The Jordan
frame requires a definite, small, negative vacuum density of virtual photons
whereas the Einstein frame requires a small positive density of {'}upwards{'}
accelerating virtual particles. The theory thereby naturally connects
gravitational theory with quantum expectations of the vacuum. These virtual
densities are coupled to curvature and approach zero simultaneously with it.
Thus it seems that SCC predicts a limit to the Casimir force as a function
of space-time curvature that may be detectable. A rough calculation,
dependent on the sensitivity of the apparatus, indicates that such a
detection may be made in the solar system somewhere between the orbits of
Jupiter and Saturn.

The third type test is being performed at present; it is the Gravity Probe B
measurement of geodetic precession of a gyroscope in polar orbit around the
Earth. On the one hand, the {'}frame dragging{'} prediction of that
experiment is given by the expression: 
\begin{equation}
\overset{3}{g}_{i0}=-4G_{m}\left( \frac{1+\gamma _{r}}{2}\right) \int \frac{%
\overset{1}{T}^{i0}(x^{\prime },t)}{\left\vert x-x^{\prime }\right\vert }%
d^{3}x\text{,}  \label{54}
\end{equation}%
so that the SCC values, $\gamma _{r}=\frac{1}{3}$ and $G_{m}=\frac{3}{2}%
G_{N} $, give the same result as the GR values $G_{m}=G_{N}$ and $\gamma
_{r}=1$. On the other hand, the geodetic precession is given by the
expression 
\begin{equation}
\frac{1}{2}\left( 2\gamma _{r}+1\right) \frac{G_{m}M_{\oplus }}{R^{3}}%
\mathbf{v}_{s}\times \mathbf{X}\text{ ,}  \label{55}
\end{equation}%
where $M_{\oplus }$ is the mass of the Earth; so that the SCC prediction is $%
\frac{5}{6}$ of that of the GR prediction of 6.6144 arc sec/yr about a
direction perpendicular to the plane of the orbit. In SCC there is a Thomas 
precession, which has to be subtracted from the geodetic precession, of 
 \begin{equation}
\frac{1}{6} \frac{G_{m}M_{\oplus }}{R^{3}}%
\mathbf{v}_{s}\times \mathbf{X}\text{ ,}  \label{56}
\end{equation}%
Therefore, the SCC theory prediction of a N-S precession of the GP-B gyroscope 
is $%
\frac{2}{3}$ of that of the GR prediction, or just 4.4096 arcsec/yr. 
The SCC expectation of this experiment is that if the results are interpreted 
within a GR environment (setting $G_m = G_N$) then the values obtained for 
$\gamma _r$ will be grossly inconsistent. If SCC is correct then such a GR analysis 
would yield $\gamma _r= 1$ from the frame dragging experiment but $\gamma _r = 0.5$ 
from geodetic precession. Such a gross inconsistency would be evidence falsifying 
the equivalence principle. This crucial measurement will be the first experiment 
that is able to distinguish between the two theories.

\subsection{Potential Problems and Tests of the EEP}

SCC contains a gross violation of the equivalence principle, which makes it
problematical to believe that the theory could be concordant with the
laboratory experimental tests of the EEP. There are two questions to answer.
The first is, "If there are many different types of matter present, how does
the scalar field $\phi$ couple to the individual matter components? In
particular, how is the lack of conservation of the total stress-energy $%
T_{M\mu\nu}$ shared among the different fields?" The second question, the
answer of which depends partly on the first and partly on first principles,
is, "If it is really true that {'}photons fall at a faster rate than
particles by a factor of 3/2{'}, then electromagnetic fields must couple
very differently to the metric and/or $\phi$ than other forms of matter.
However, as a nontrivial portion of the mass-energy of atoms is
electromagnetic in origin, and this fraction varies substantially from
material to material, would not one expect different types of materials to
fall at different rates?"

In answer to the first question we recall that our scalar and creation field
equations are given by 
\begin{equation*}
\Box \phi =4\pi T_{M}
\end{equation*}
and%
\begin{equation*}
\nabla _{\mu }T_{M}^{\mu }{}_{\nu }=\frac{1}{2}\frac{\nabla _{\nu }\phi }{%
\phi }T_{M}
\end{equation*}
respectively, so the scalar field and the 'lack of conservation' couple to
different matter components according to their trace $T_{M}$.

The second question requires a little more thought. One of the basic
principles of the theory is that of Mach. This is enshrined by the scalar
field equation. As totally relativistic forms of energy, which have an
equation of state $p = \frac 13 \rho c^2$, are traceless they are decoupled
from the scalar field. This is entirely consistent with Mach's principle and
Special Relativity; for as according to the latter the speed of light is
invariant across all frames of reference then one cannot define an inertial
frame of reference using the distribution of light and other forms of free
relativistic energy within the universe. This is the macroscopic case.

However the microscopic case is more uncertain as it is encompasses the
interface between gravitational and quantum physics. We note that the
similar question of how the gravitational and scalar fields couple to matter
on the atomic and nuclear scale has not yet been answered in the GR or BD
theories either. What is clear is that, in order for SCC not to contradict
the experimental results of the EEP, it has to be assumed that once
electromagnetic energy is atomically bound and hence {'}located{'} within an
atom, so its mass equivalent must be accounted for with that atom as far as
Mach's principle is concerned. Such energy does therefore contribute to the
overall trace of the matter it is bound to. The density and pressure values
that enter into the set of field equations must therefore be the average
macroscopic values of the continuum of matter. The question is whether or
not this requirement is plausible.

Accordingly, how then do different materials accelerate within a
gravitational field? Present day tests of the EEP have confirmed that
different elements such as gold and aluminium fall at the same rate to
within one part in $10^{-14}$. Treating different elements as perfect
fluids, the violation of the EEP is due to the pressure of stress energy
compared with rest mass. The full equation of motion is given by [21] 
\begin{eqnarray}
\frac{d^{2}r}{dt^{2}} &=&-\frac{3}{2}\left[ 1-\frac{\left( \rho
c^{2}-3p\right) }{3\rho c^{2}}\right] \frac{G_{N}M_{\oplus }}{r^{2}}
\label{57} \\
&&+\left[ 3-2\frac{\left( \rho c^{2}-3p\right) }{\rho c^{2}}\right] \left( 
\frac{G_{N}M_{\oplus }}{rc^{2}}\right) \frac{G_{N}M_{\oplus }}{r^{2}}+%
\mathit{O}\left( \frac{G_{N}M_{\oplus }}{rc^{2}}\right) ^{2}\text{.}  \notag
\end{eqnarray}%
Deviations from the EEP are thus one part in $\frac{3p}{2\rho c^{2}}$ and as
the second term is a factor $7\times 10^{-10}$ smaller than the first, on
the Earth's surface, it can be ignored. The deviation for aluminium under
atmospheric pressure is therefore one part in $6\times 10^{-16}$. However
experiments to such accuracy are actually carried out in a vacuum in which
the only pressure is due to the internal stress set up by the weight of the
object. The maximum internal pressure of a body of height $h$ is $\rho gh$,
therefore the deviation from the EEP is one part in $ghc^{-2,}$ or about $%
h\times 10^{-16}$ where h is measured in metres. Consequently in a typical E%
\"{o}tvos-type experiment the violation of the EEP would be about one part
in $10^{-17}$ or about three orders of magnitude smaller than the present
day sensitivity of the experiment.

\section{The Cosmological Solution}

Using the cosmological principle the usual assumptions of homogeneity and
isotropy can be made to obtain the cosmological solutions to the field
equations, [3]. The privileged CoM frame in which physical units may be
defined for any epoch is now the {'}rest frame{'} for the universe as a
whole in which the Cosmic Background Radiation is isotropic. Transformations
between the Jordan, and Einstein, frames have to be made as appropriate.

The gravitational field equation is the same as the BD equation with w =
-3/2; where a superimposed dot signifies the time derivative, 
\begin{equation}
\left( \frac{\overset{.}{R}}{R}\right) ^{2}+\frac{k}{R^{2}}=+\frac{8\pi \rho 
}{3\phi }-\frac{\overset{.}{\phi }\overset{.}{R}}{\phi R}-\frac{1}{4}\left( 
\frac{\overset{.}{\phi }}{\phi }\right) ^{2}\text{ ,}  \label{58}
\end{equation}%
which is obtained from the time-time and space-space components. The second
equation derived from the gravitational field equation also includes the
scalar field equation, 
\begin{equation}
\frac{\overset{..}{R}}{R}+\left( \frac{\overset{.}{R}}{R}\right) ^{2}+\frac{k%
}{R^{2}}=-\frac{1}{6}\left( \frac{\overset{..}{\phi }}{\phi }+3\frac{\overset%
{.}{\phi }\overset{.}{R}}{\phi R}\right) +\frac{1}{4}\left( \frac{\overset{.}%
{\phi }}{\phi }\right) ^{2}\text{ .}  \label{59}
\end{equation}%
The third equation is the same as the BD scalar field equation: 
\begin{equation}
\overset{..}{\phi }+\,3\frac{\overset{.}{\phi }\overset{.}{R}}{R}=4\pi
\left( \rho -3p\right) \text{ .}  \label{60}
\end{equation}%
In GR and BD a fourth equation is obtained from the conservation
requirement: 
\begin{equation}
\overset{.}{\rho }\,=-3\frac{\overset{.}{R}}{R}\left( \rho +p\right) \text{ ,%
}  \label{61}
\end{equation}%
but here it is replaced by, 
\begin{equation}
\overset{.}{\rho }\,=-3\frac{\overset{.}{R}}{R}\left( \rho +p\right) +\frac{1%
}{8\pi }\frac{\overset{.}{\phi }}{\phi }\left( \overset{..}{\phi }+\,3\frac{%
\overset{.}{\phi }\overset{.}{R}}{R}\right) \text{ ,}  \label{62}
\end{equation}%
with an extra term representing cosmological {'}self-creation{'}. It is a
moot point whether the scalar field $\phi $ is generated by the distribution
of mass and or whether mass is generated by the scalar field. A fifth
equation is obtained from some equation of state, $p=\sigma \rho $, where,
for example, $\sigma =+1/3$ in a radiation dominated, and $\sigma =0$ in a
dust filled, universe, but the SCC field equations demand an exotic equation
of state. There are therefore five independent equations to solve for six
unknowns $R(t)$, $\phi (t)$, $\rho (t)$, and $\sigma $, and a sixth
relationship is provided by Stephan{'}s Law and the conservation of a free
photon{'}s energy in the Jordan frame. The boundary conditions at $t=t_{0}$
(the present epoch), are known, $R_{0}$, $\phi _{0}$, $\rho _{0}$, and $%
p_{0} $. The cosmological equations can be reduced to a relationship between 
$\rho (t)$, $R(t)$, and $\phi (t)$, 
\begin{equation}
\rho =\rho _{0}\left( \frac{R}{R_{0}}\right) ^{-3\left( 1+\sigma \right)
}\left( \frac{\phi }{\phi _{0}}\right) ^{\frac{1}{2}\left( 1-3\sigma \right)
}\text{ ,}  \label{63}
\end{equation}%
which again is the equivalent GR expression with the addition of the last
factor representing cosmological {'}self-creation{'}.

For a photon gas $\sigma =+\frac{1}{3}$ so the last expression reduces to
its GR equivalent, $\rho _{em}\propto R^{-4}$ , consistent with the PMI
principle that there is no interaction between a photon and the scalar
field. Furthermore, by Stephan{'}s law, $\rho _{em}\propto T_{em}^{4}$,
where $T_{em}$ is the Black Body temperature of the radiation, therefore the
adiabatic GR relationship 
\begin{equation}
T_{em}\propto R^{-1}  \label{64}
\end{equation}%
still holds. As the wavelength $\lambda _{em}^{max}$ at maximum intensity of
the Black Body radiation is given by 
\begin{equation}
\lambda _{em}^{max}\propto T_{em}^{-1}\text{ ,}  \label{65}
\end{equation}%
SCC retains the GR relationship $\lambda _{em}^{max}\propto R$. However the
SCC contention is that gravitational, and hence cosmological, equations have
to be solved in the Jordan frame in which $\lambda $ is invariant, thus in
this frame $R$ must be invariant as well. In other words the universe is
stationary because a co-expanding "light-ruler" would be unable to detect an
expanding universe! In the Jordan frame with 
\begin{equation}
R=R_{0}  \label{66}
\end{equation}%
the cosmological equations reduce to 
\begin{equation}
\left( 5-3\sigma \right) \frac{\overset{..}{\phi }}{\phi }=3\left( 1-3\sigma
\right) \left( \frac{\overset{.}{\phi }}{\phi }\right) ^{2}\text{ ,}
\label{67}
\end{equation}%
which has the two possible solutions; case 1, when $\sigma \neq -\frac{1}{3}$
- then 
\begin{equation}
\phi =\phi _{0}\left( \frac{t}{t_{0}}\right) ^{\frac{\left( 5-3\sigma
\right) }{2\left( 1+3\sigma \right) }}\text{ ,}  \label{68}
\end{equation}%
and case 2, when $\sigma =-\frac{1}{3}$ then 
\begin{equation}
\phi =\phi _{0}\exp \left[ H(t-t_{0})\right] \text{ ,}  \label{69}
\end{equation}%
where H is some as yet undetermined constant of dimension $\left[ T\right]
^{-1}$.

Case 1 is that of a universe empty except for a false vacuum, in which the
presence of any other energy or matter forces the solution to take on case 2
in which the cosmological density is given by 
\begin{equation}
\frac{8\pi \rho }{\phi _{0}}=H^{2}\exp \left[ H\left( t-t_{0}\right) \right] 
\text{ .}  \label{70}
\end{equation}%
Assuming baryon conservation in a stationary universe, the inertial mass of
a fundamental particle, $m_{i}$, is given by 
\begin{equation}
m_{i}=m_{0}exp\left[ H\left( t-t_{0}\right) \right] \text{ .}  \label{71}
\end{equation}%
Cosmological red shift is not interpreted as an effect caused by
cosmological expansion, but rather as gain in the mass of the apparatus
measuring it, as with gravitational red shift. Observations of the
cosmological red shift identify $H$ as Hubble{'}s {'}constant{'}.

It is now necessary to transform from the Jordan frame used in the theory
into the system of measurement used in the laboratory, that is the Einstein
frame. The secular cosmological gain in inertial mass causes atoms and
therefore rulers to {'}shrink{'} and atomic clocks to {'}speed up{'} when
compared to their Jordan frame equivalents. Care has to be taken in
interpreting cosmological measurements, as they are often observations of a
mixture of Jordan and Einstein frame effects. In the Jordan frame distance
and time are given by $r$, $R$ and $t$; in the Einstein frame they shall be
expressed by italicised $\mathit{r}$, $\mathit{R}$ and $\mathit{t}$. In the
Jordan frame if the origin of the time system is defined to be the present
moment, $t_{0}=0$, then as $t\rightarrow -\infty $ the masses of fundamental
particles tend towards zero, and hence the sizes of the particles will tend
towards infinity; whereas in the Einstein frame at this 'Big Bang' epoch,
the universe has zero volume although particles are of normal size. In
either case the universe is equally crowded! Hence it is natural in the
Einstein frame to define this epoch, marking the initial moment of the 'Big
Bang' to be the origin, $\mathit{t}=0$. The two time systems then relate
together as follows: 
\begin{equation}
\mathit{t}=H^{-1}exp\left( Ht\right) \text{ and }t=H^{-1}ln\left( H\mathit{t}%
\right) \text{ .}  \label{72}
\end{equation}

In the Einstein frame, $\mathit{R = R_0 (t/t_0)}$, and the universe thus
appears to expand linearly from a {'}big bang{'}. Because the deceleration
parameter $q = - \left(\mathit{\overset{..}{R}}/H^2\mathit{R}\right)$ equals
zero in SCC as $\mathit{\overset{..}{R}} = 0$, the horizon, smoothness and
density problems of classical GR cosmology, which all arise from a positive
non-zero q, do not feature in SCC. Therefore in this theory it is
unnecessary to invoke Inflation theory and indeed, instead, SCC might be
considered to be a form of 'Continuous Inflation'.

The above equations give the cosmological density as 
\begin{equation}
\rho _{0}=\frac{H_{0}^{2}}{8\pi G_{N}}\text{ ,}  \label{73}
\end{equation}%
where $G_{N}$ is the value of $\phi ^{-1}$ in the present epoch, from which
the total density parameter 
\begin{equation}
\Omega =\frac{1}{3}\text{ .}  \label{74}
\end{equation}%
The cosmological equations require the cosmological pressure to be $p=-\frac{%
1}{3}\rho $. This exotic equation of state appears to have the form of a
non-zero cosmological constant. However instead, in a similar fashion to the
One Body case, there is a false vacuum component of the universe created by
the scalar field. SCC therefore suggests that there is a {'}remnant{'}
vacuum energy made up of contributions of zero-point energy from every mode
of every quantum field that would have a natural energy {'}cut-off{'} $%
E_{max}$ determined and limited by the solution to the cosmological
equations. This component of false vacuum is calculated to have a density of
one-third the total density; hence the remaining cold matter (visible and
dark) density parameter is determined by 
\begin{equation}
\Omega _{f}v=\frac{1}{9}\text{ and therefore }\Omega _{b}=\frac{2}{9}\ =0.2%
\overset{.}{2}\text{.}  \label{75}
\end{equation}%
The difference between $\Omega $ and $\Omega _{b}$ could be interpreted as
the hot dark matter component of {'}missing mass{'} or, with negative
pressure, it might have been identified with {'}dark energy{'} or {'}%
quintessence{'} [22], [23], [24].

The curvature constant, $k$, is found to be positive, so the universe is
finite and unbounded, with a {'}radius of curvature{'} 
\begin{equation}
R_{0}=\frac{\sqrt{12}}{H_{0}}c\text{ .}  \label{76}
\end{equation}

By definition the mass of a fundamental particle, $\mathit{m}$, is constant
in the Einstein frame, however when compared with the energy of a free
photon, the mass is given by, 
\begin{equation}
m=m_{0}\left( \mathit{\ \frac{t}{t_{0}}}\right) =m_{0}\left( \mathit{\frac{R%
}{R_{0}}}\right) \text{ ,}  \label{77}
\end{equation}%
which is normally interpreted in this frame as the free photon suffering a
red shift, $1+z=R_{0}/R$ . Similarly, using Einstein frame time, $\phi $ is
given by 
\begin{equation}
\phi =G_{N}^{-1}\left( \mathit{\frac{t}{t_{0}}}\right) \text{ ,}  \label{78}
\end{equation}%
but this variation is normally 'hidden' by the compensating change in atomic
masses that causes GM to be constant.

Nevertheless this relationship might explain the Large Numbers Hypothesis
(LNH) relationship $G \approx T^{-1}$ where $G$ and $T$ are the normal LNH
dimensionless values of the gravitational constant and the age of the
universe respectively.

The observation of any variation in $G$ will depend on the techniques used.
If Hubble time is of the order $15\times 10^{9}$ years then according to the
theory: 
\begin{equation}
\frac{\overset{.}{G}\left( t\right) }{G\left( t\right) }\approx
-6.10^{-11}yr^{-1}\text{ .}  \label{79}
\end{equation}%
This may have already been observed. Krasinsky's et al. result from the
analysis of the residuals of planetary longitudes [25] is: 
\begin{equation}
\frac{\overset{.}{G\left( t\right) }}{G\left( t\right) }\approx +\left( 4\pm
0.8\right) \times 10^{-11}yr^{-1}  \label{80}
\end{equation}%
with a caveat that the sign might be reversed. However they also reported
the contradictory null result of Hellings et al., [26], determined from
accurate observations of the Viking Landers and the Mariner 9 spacecraft
with the effect that a possible falsification of GR was not followed up.
However SCC may explain the discrepancy between these two results. The
residuals of planetary longitudes are {'}remote{'}, gravitational,
observations that are to be interpreted in the Jordan frame, whereas the
radar ranging of a spacecraft depended on an atomic clock. In which case the
secular increase in atomic mass would affect clock rate and hence compensate
for the variation in G to give a null result.

The gravitational field of a massive body remains invariant over
cosmological time in the Einstein frame. This effect, which manifests itself
as a {'}time slip{'} between atomic and gravitational clocks, that is,
between {'}atomic time{'} and {'}ephemeris time{'}, also explains the
anomalous sunward acceleration observed on the Pioneer spacecraft [4], [5],
[27].

Summing up, in the JF(E), where energy is conserved but energy-momentum is
not, photons are the means of measuring length, time and mass. Proper mass
increases with gravitational potential energy and therefore cosmological red
shift is caused by a secular, exponential, increase of particle masses and
not cosmological expansion. The universe is stationary, in which atomic
rulers {'}shrink{'} exponentially, and eternal, in which atomic clocks {'}%
speed up{'} exponentially.

\subsection{An explanation for some anomalous observations}

The consequence of this theory is the realisation that there are two
distinct ways of interpreting observations of the universe. In a laboratory
on Earth scientific observations that define units of length, time and
mass/energy have to be referred to an atomic standard. However,
astrophysical and cosmological observations only sample photons from the
depths of the universe. How then does the measurement of the standard units
made in a laboratory here and now on Earth relate to an event that occurred
millennia ago in a distant part of the universe? In particular, the problem
is rooted in the red shift of photons over and above that caused by the
Doppler effect.

Because of the equivalence principle GR defines the proper rest mass of a
particle to be invariant, therefore that theory requires the measures of
standard units to be atomic {'}rigid{'} rulers and atomic {'}regular{'}
clocks. However, the penalty for doing so is to violate the conservation of
mass-energy as described in the Introduction above. On the other hand, if a
gravitational theory were to locally conserve energy, as in the theory of
SCC, which for consistency also subsumes Mach's Principle, then atomic rest
masses would vary with gravitational potential energy. If this indeed occurs
then a choice may be made as to the invariant standard by which units of
length, time and mass/energy are measured. This choice of the standard for
comparison is between a {'}standard{'} atom, taken from a laboratory, or a {'%
}standard{'} photon, sampled from the CMB. Observations of the cosmos would
then reveal one of two possible universes: either a stationary universe that
is eternal with no origin in time, the JF(E) interpretation, or a strictly
linearly expanding or {'}freely floating{'} universe that has had an {'}%
origin{'} in a {'}Big Bang{'} at one {'}Hubble Time{'} in the past, the EF
interpretation. Either model would be a valid interpretation of the data,
the JF(E) would be the appropriate frame to observe gravitational orbits and
the curvature of space-time, and the EF would be the appropriate frame to
observe atomic processes such as primordial nucleo-synthesis.

It is remarkable that both these models, the stationary universe and the
freely coasting universe, have already been independently investigated and
both have been found to be surprisingly concordant with accepted
cosmological constraints, including the Big Bang nucleo-synthesis
abundances, distant Type Ia supernovae observations and the WMAP CMB
anisotropy data.

Ostermann, [5], investigated the stationary universe heuristically, he
included an exponential cosmological time slip in his theory and found it
was able to explain the Pioneer anomaly and fit the standard concordance
model perfectly [28].

The strictly linearly expanding or freely coasting model has been
investigated by Kolb, [29], Dev, Safonova, Jain and Lohiya, [30], Gehaut,
Mukherjee, Mahajan and Lohiya, [31], and Gehaut, Kumar, Geetanjali and
Lohiya, [32]. Their motivation in exploring such a cosmology was the
recognition that such a model would not have suffered from the original
density, smoothness and horizon problems of the standard GR theory. The
latter paper reviews their results and finds the freely coasting universe
fits the Type 1a supernovae data. Moreover, the recombination history gives
the location of the primary acoustic peaks of the WMAP data in the same
range of angles as that given in standard cosmology. A further remarkable
result is their analysis of nucleo-synthesis in the Big Bang. They calculate
that a baryon entropy ratio of $\eta =5\times 10^{-9}$ yields 23.9 per cent
Helium and $10^{8}$ times the metallicity of the standard scenario, which is
still of the same order of magnitude as seen in the lowest metallicity
objects. Therefore, one prediction of the theory is that a significant
proportion of intergalactic medium metallicity, observed from the Lyman a
forest of distant quasar spectra, should be primordial.

A further consequence is, interestingly, that the production of this amount
of helium requires a baryon density parameter of about 0.2. As the total
non-false vacuum energy density is required by SCC to be only 0.22, there is
no need for unknown dark matter. In SCC, this component of the cosmic
density parameter is in the form of intergalactic cold baryonic matter.

Furthermore, the cosmological solution requires the universe to have an
overall density parameter of only one third, yet be closed and conformally
spatially flat. Hence, the theory does not require dark energy, or a
significant amount of dark matter, to account for the present cosmological
constraints.

In SCC the vacuum energy density is negative and stably determined by the
field equations, with no cosmological constant thus fulfilling the
requirements of present superstring models. The high baryonic matter density
also relieves the problem over the galaxy mass profiles that require low
energy density to fit the observed cusp.

Another observable effect arises in the JF(E) as a result of the variation
in $m(t)$. If angular momentum is conserved then $mr^{2}\omega $ is
constant. An atomic radius is inversely proportional to its mass, so if the
mass increases secularly, the radius will shrink. 
\begin{equation}
\text{If }m(t)=m_{0}exp\left( H_{0}t\right) \text{, then }%
r(t)=r_{0}exp\left( -H_{0}t\right)  \label{81}
\end{equation}%
\begin{equation}
\text{and if }\frac{d}{dt}\left( mr^{2}\omega \right) =0  \label{82}
\end{equation}%
\begin{equation}
\text{Then }\frac{\overset{.}{\omega }}{\omega }=-\left( \frac{\overset{.}{m}%
}{m}+2\frac{\overset{.}{r}}{r}\right) =+H_{0}\text{ ,}  \label{83}
\end{equation}%
and solid bodies such as the Earth should spin up when measured by JF(E),
(ephemeris) time. It has indeed been reported that this is observed. As
mentioned above, the review by Leslie Morrison and Richard Stephenson, [6],
[7], studying the analysis of the length of the day from ancient eclipse
records reported that in addition to the tidal contribution there is a
long-term component acting to decrease the length of the day which equals 
\begin{equation}
\triangle T/day/cy=-6\times 10^{-4}\text{ sec/day/cy.}  \label{84}
\end{equation}%
This value, equivalent to $H=67km.sec^{-1}/Mpc$, is remarkably close to the
best estimates of $H_{0}$. The lunar orbit, being a geodesic through
space-time, makes a JF(E) clock, it "tells ephemeris time", so this
observation is exactly that predicted by SCC. However at least part of this
spin up is probably caused by a decrease of the Earth's moment of inertia.

\section{A novel representation of space-time geometry}

In the stationary model of the SCC JF(E), a space-like slice of space-time
produces a hyper-sphere independent of the slice chosen, whereas a time-like
slice produces a cylinder. It is Einstein's stationary cylindrical model. In
the EF, on the other hand, consecutive space-like slices produce a series of
hyper-spheres whose radii are proportional to the age of the space slice
chosen, whereas a time-like slice produces a conical model. The series of
hyper-spheres that steadily increase may be re-arranged into a series of
concentric hyper-spheres. The model is now that of a radial time universe.
Treating time as a radial coordinate has several attractions. As a radial
coordinate it has an origin but no negative values, time just did not exist
before $t = 0$, the Big Bang. A child-like question may be, "What is the
universe expanding into?" The answer may be given, "The future". In such a
model the expansion of the universe and the passing of time may be seen as
two different experiences of the same phenomenon, yet what that phenomenon
may be still remains a mystery.

\section{Conclusions}

Whereas GR cosmology can be considered as a model with two free parameters, $%
\Omega $ and $H$, this theory, SCC, has only one, $H$, and its predictions
are highly determined. Nevertheless in local experiments and in the basic
cosmological parameters the theory does seem to yield reasonable predictions
consistent with experiment. The theory explains the present quandary about
the observed distant supernovae observations and yields predicted density
parameters consistent with observation, which otherwise might be interpreted
as the effects of a cosmological constant and/or quintessence. Furthermore,
as a new theory of gravitation, SCC is readily open to falsification in the
definitive experiments described earlier. The theory might be described as {'%
}fully Machian{'} as 
\begin{equation}
\Omega _{b}=\frac{2}{9}\text{ yields }G_{N}=\frac{H_{0}^{2}}{12\pi \rho _{b}}%
\text{ ,}  \label{85}
\end{equation}%
and therefore the gravitational constant is fully determined by a knowledge
of the large scale structure of the universe. It remains to be seen whether
nucleo-synthesis and gravitational instability analysis in the SCC universe
confirm the reasonable values for primordial element relative abundance and
yields a plausible model for observed matter distribution.

Two of the attractive features of this theory are those of the self-creating
nature of the universe ex nihilio, and the transposition of the initial
moment of EF time, $t = 0$, back into the infinite past of JF(E) time, $t
\rightarrow -\infty$ . Inertial rest mass is seen as a measure of total
potential energy, that is the energy required to {'}lift{'} a particle out
of the {'}Big Bang{'} to the present day. Inertial mass is created out of
the zero point energy field by the self-contained gravitational and scalar
fields within the universe, hence the suggested title of the theory. Finally
there is a choice in the way time may be measured; depending on the clock
used to measure it, the universe is understood either to have had a {'}
beginning{'}, or to be eternal. The latter case thereby avoids philosophical
problems associated with an origin.

Correspondence should be addressed to:

\qquad \qquad G.A.B. (e-mail: garth.barber@virgin.net)


\section{References}

\begin{thebibliography}{99}
\bibitem{1} Bennett, C. et al.: Feb 2003, astro-ph/0302207. First Year
Wilkinson Microwave Anisotropy Probe (WMAP) Observations: Preliminary Maps
and Basic Results.

\bibitem{2} Hinshaw, G. et al.: Feb 2003, astro-ph/0302217. First Year
Wilkinson Microwave Anisotropy Probe (WMAP) Observations: Angular Power
Spectrum.

\bibitem{3} Barber, G.A.: 2002, Astrophysics and Space Science 282, 4, pp
683-731. A New Self Creation Cosmology.

\bibitem{4} Anderson, J.D. \& Laing, P.A. \& Lau, E.L. \& Liu, A.S. \&
Nieto, M.M. \& Turyshev, S.G.: Apr 2002, Physical Review D, 65, 8, id.
082004. arXiv:gr-qc/0104064 v4 11. Study of the anomalous acceleration of
Pioneer 10 and 11.

\bibitem{5} Ostermann, P.: Dec 2002, arXiv:gr-qc/0212004. Relativity Theory
and a Real Pioneer Effect.

\bibitem{6} Morrison, L. \& Stephenson, F.R.:1998, Astronomy \& Geophysics
Vol. 39 October. The Sands of Time and the Earth's Rotation

\bibitem{7} Stephenson, F.R.:2003, Astronomy \& Geophysics Vol. 44 April.
Historical eclipses and Earth's rotation.

\bibitem{8} Krasinsky et al.: 1985, I.A.U. Symposium No.114; Relativity in
Celestial Mechanics and Astrometry. Relativistic effects from planetary and
lunar observations of the XVIII-XX Centuries. 1985.

\bibitem{9} Barber, G.A.:1982, Gen Relativ Gravit. 14,117. On Two Self
Creation Cosmologies.

\bibitem{10} Brans, C.H. \& Dicke, R.H.: 1961, Phys. Rev. 124, 925. Mach's
Principle and a Relativistic Theory of Gravitation.

\bibitem{11} Barber, G.A.: 2004, astro-ph/0401136. The Self Creation
challenge to the cosmological 'concordance model'.

\bibitem{12} Brans, C.H.: 1987, Gen Relativ Gravit. 19, 949. Consistency of
field equations in self-creation cosmologies,

\bibitem{13} Barber, G.A.: 2003b, arXiv:gr-qc/0302088. The derivation of the
coupling constant in the new Self Creation Cosmology.

\bibitem{14} Milne, E.A.: 1935, Clarendon Press: Oxford. Relativity,
Gravitation and World Structure.

\bibitem{15} Milne, E.A.: 1948, Clarendon Press: Oxford. Kinematic
Relativity - a sequel to Relativity, Gravitation and World Structure.

\bibitem{16} Noether, E.: 1918, Nachr. d. K\"{o}nig. Gesellsch. d. Wiss. zu G%
\"{o}ttingen, Math-phys. Klasse.

\bibitem{17} Byers, N.: 1998, Proceedings of a Symposium on the Heritage of
Emmy Noether (1996). arXiv:physics/9807044. E. Noether's Discovery of a Deep
Connection Between Symmetries and Conservation Laws

\bibitem{18} Weyl, H.: 1918, 'Gravitation und Electriticitat'
Sitzungsberichte der Preussichen Akad. d. Wissenschaften, English
translation, 1923, in: The Principle of Relativity, Dover Publications.

\bibitem{19} Butterfield, J. \& Isham, C. J.: 2001, Physics meets Philosophy
at the Planck Scale, ed. by C. Callender and N. Huggett. Cambridge
University Press.

\bibitem{20} Misner, Thorne and Wheeler: 1973, Gravitation. (Equation
40.15). Freeman.

\bibitem{21} Barber, G.A.: 2002, Astrophysics and Space Science 282, 4, pp
683-731. A New Self Creation Cosmology , page 706 derived from Equation 126.

\bibitem{22} Cruz, N., et al.: 1998, Phys. Rev. D 5812, 12. Art. no. 123504,
Matter scalar field in a closed universe.

\bibitem{23} Huey, G., et al.: 1999, Phys.Rev. D 5906, 6. - Art. no. 063005,
Resolving the cosmological missing energy problem.

\bibitem{24} Zlatev, I., et al.: 1999, Phys. Rev. Lett. 82 (5), 896-899.
Quintessence, cosmic coincidence, and the cosmological constant.

\bibitem{25} Krasinsky et al.: 1985, I.A.U. Symposium No.114; Relativity in
Celestial Mechanics and Astrometry. Relativistic effects from planetary and
lunar observations of the XVIII-XX Centuries.

\bibitem{26} Hellings, R.W. et al.: 1983, Phys. Rev. Letts., 51 pg. 1609.

\bibitem{27} Mbelek, J.P. \& Michalski, M.: 2003, arXiv:gr-qc/0310088. Can
conventional forces really explain the anomalous acceleration of Pioneer
10/11?

\bibitem{28} Ostermann, P.: 2003, arXiv:astro-ph/0312655. The Concordance
Model - a Heuristic Approach from a Stationary Universe.

\bibitem{29} Kolb, E.W.: 1989, The Astrophysical Journal 344, 543. A
coasting cosmology.

\bibitem{30} Dev, A. \& Safonova, M. \& Jain, D. \& Lohiya, D.: 2002,
arXiv:astro-ph/0204150. Cosmological Tests for a Linear Coasting Cosmology.

\bibitem{31} Gehaut, S. \& Mukherjee, A. \& Mahajan, S. \& Lohiya, D.: 2002,
arXiv:astro-ph/0209209. A "Freely Coasting" Universe.

\bibitem{32} Gehaut, S. \& Kumar, P. \& Geetanjali. \& Lohiya, D.:
2003,arXiv:astro-ph/0306448. A Concordant "Freely Coasting Cosmology".
\end{thebibliography}
\end{document}